# Ferromagnetic resonance spin pumping in CoFeB with highly resistive non-magnetic electrodes.

Dong-Jun Kim[1], Sang-Il Kim[2], Seung-Young Park[2], Kyeong-Dong Lee[1], and Byong-Guk Park[1,a)]

[1] Department of Materials Science and Engineering, KAIST Institute for the Nanocentury, KAIST, Daejeon 305-701, Republic of Korea

[2] Division of Materials Science, KBSI, Daejeon 305-806, Republic of Korea

a) Corresponding Authors:

Prof. Byong-Guk Park

Department of Material Science and Engineering, KAIST

Email: bgpark@kaist.ac.kr

Tel.: +82-42-350-3330, Fax: +82-42-350-3310.

## Abstract

The relative contribution of spin pumping and spin rectification from the ferromagnetic resonance of CoFeB/non-magnetic bilayers was investigated as a function of non-magnetic electrode resistance. Samples with highly resistive electrodes of Ta or Ti exhibit a stronger spin rectification signal, which may result in over-(or under-)estimation of the spin Hall angle of the materials, while those with low resistive electrodes of Pt or Pd show the domination of the inverse spin Hall effect from spin pumping. By comparison with samples of single FM layer and an inverted structure, we provide a proper analysis method to extract spin pumping contribution.

The spin Hall effect (SHE), which converts charge current into spin accumulation[1-4], can be utilized in various types of spintronic devices, such as current control of magnetization orientation,[5,6] high-speed domain wall motion,[7,8] generation of spin dynamics,[6,9] and detection of spin transport in non-magnetic materials,[10] etc. The SHE is quantified by the spin Hall angle ($\theta_{SH}$) representing the effectiveness of the charge-spin conversion,[11] which is a material property related to spin-orbit coupling strength. Therefore, the quantification of the $\theta_{SH}$ is of great importance in order to understand this effect and to apply the SHE to novel spintronic devices.

Spin pumping from the ferromagnetic resonance (FMR) of ferromagnetic/non-magnetic (FM/NM) bilayers is a widely used method to determine the $\theta_{SH}$.[12,13] The spin current generated by FMR is injected into an NM electrode where dc voltage is induced as a result of inverse SHE.[14,15] By measuring the dc voltage of the bilayers, the $\theta_{SH}$ of an NM material can be obtained. However, in similar bilayer structures, dc voltage can be also comprised of spin rectification, which originates from the anisotropic magnetoresistance (AMR) or anomalous Hall effect (AHE) of the FM layer.[16] The spin rectification effect may contaminate the spin pumping signal, so the separation of the spin rectification effect is inevitably required to obtain the $\theta_{SH}$.[12,17,18] This may be one of the reasons for the inconsistent $\theta_{SH}$ values that are experimentally reported for the nominally identical materials.[19-21] The contributions of spin pumping and spin rectification are normally distinguished by their symmetries. The spin pumping is largest at a resonance magnetic field where a Lorentzian symmetric dc voltage is expected. On the contrary, the spin rectification originated from a non-zero electric field in a sample gives rise to an asymmetric voltage at the resonance field due to a phase difference of $\pi/2$ between electric and magnetic fields.[12] Since induced current by the electric field depends on the resistance of a material and the spin rectification is only due to the current flowing in the FM layer, a resistance ratio between the FM and NM layers determines the relative contribution of the spin rectification to the total dc voltage. In this work, we have investigated the spin pumping and the spin rectification in CoFeB/NM bilayers with different resistances using an FMR in a microwave cavity resonator. For samples with low resistive NM electrodes of Pt or Pd, the spin pumping effect dominates because induced current mostly shunts through the NM layer.[22] On the other hand, the samples with highly resistive electrodes of Ta or Ti exhibit stronger spin rectification voltage. In order to extract a pure spin pumping effect, we measured the FMR of a single CoFeB layer and subtracted the contribution of the FM layer from the total dc voltage. Moreover, the results are confirmed by comparison to samples with inverted structures.

Samples of $Co_{32}Fe_{48}B_{20}$(10nm)/NM(10nm) bilayer structures were deposited by magnetron sputtering on thermally-oxidized Si substrates, where NM is Pt, Pd, Ta, and Ti whose resistivities are 23, 31, 212, and 308 $\mu\Omega\cdot$cm, respectively. Note that the resistivity of a single CoFeB film is 178 $\mu\Omega\cdot$cm. A 1×2 mm sample was placed in a circular microwave cavity (TE011 mode) in a nodal position where the magnetic (electric) field is maximum (minimum). The external magnetic field ($H$) was applied to the in-plane direction normal to the electrode's contact, as shown in Figure 1(a). The microwave frequency was 9.46 GHz, and the power was 50 mW. The resonance field ($H_0$) was around 857 Oe. With this configuration, we investigated FMR absorptions and dc voltages of the bilayers using a Lock-in amplifier (SR830) and a nanovoltmeter (Agilent 34420A), respectively.

Figure 2 shows measured dc voltages of the samples with different NM electrodes at a resonant magnetic field where the FMR absorption is maximized (not shown here). The signals were analyzed to distinguish symmetric ($V_{sym}$) and asymmetric ($V_{asym}$) components using the equation below, and the magnitude and/or sign of those components varies depending on NM layer materials.

$$V = V_{sym} \frac{\Delta H^2}{(H-H_0)+\Delta H^2} + V_{asym} \frac{\Delta H(H-H_0)}{(H-H_0)+\Delta H^2} \quad (1)$$

Here, the $\Delta H$ is an FMR line width extracted from a Lorentzian absorption curve, as shown in Figure 1 (b). We firstly observed a negative sign of the $V_{sym}$ for a sample with Ta, but a positive sign for a Pd or Pt sample. This can be understood by the fact that the $\theta_{SH}$ of Ta is an opposite sign to that of Pt.[6] Secondly, the magnitude of the $V_{asym}$ strongly depends on the resistivity of the NM materials. The higher resistive materials show a larger contribution of the $V_{asym}$. The $V_{asym}$ is known to originate from the spin rectification of the FM layer, possibly due to the AMR or AHE effect. Note that the cavity resonator in our experiment is specially designed to minimize the electric field at the sample location, but not to completely suppress it, as evidenced by the existence of the $V_{asym}$. In order to properly separate the pure spin pumping contribution where the $\theta_{SH}$ is extracted, an identical experiment was performed for a single FM CoFeB sample, as shown in Figure 1(c). There is no NM layer where the spin current generated by FMR spin pumping is injected; thus, no $V_{sym}$ is expected if the $V_{sym}$ is solely originated from spin pumping. However, the single CoFeB layer shows both $V_{sym}$ and $V_{asym}$ components of the dc voltage, and their magnitude ratio of $(V_{sym}/V_{asym})_{CoFeB}$ is ~0.86. This is inconsistent to purely asymmetric dc voltage obtained from a single NiFe sample.[12] The symmetric contribution may be attributed to the existence of electric field component which is not $\pi/2$ phase difference from magnetic field,[17] but a further study is still needed to clarify the origin of the $V_{sym}$ in the single CoFeB. Nevertheless, these results demonstrate that the effect should be taken into consideration when the $\theta_{SH}$ is extracted, and that the simply subtracting the $V_{asym}$ from the total dc voltage may result in incorrect estimation of the $\theta_{SH}$, as the $V_{sym}$ of the FM/NM bilayer are contributed not only by spin pumping, but also by spin rectification of the FM CoFeB layer.

By assuming that the contribution of CoFeB layer is identical for each sample, we can correct the measured $V_{sym}$ with the following expression:

$$V_{corr\,sym} = V_{sym} - V_{asym} \times \left(\frac{V_{sym}}{V_{asym}}\right)_{CoFeB}, \qquad (2)$$

where $V_{corr\,sym}$ is a corrected dc voltage solely from spin pumping and the second term of the right-hand side of the equation represents the contribution of spin rectification to $V_{sym}$. The results plotted in Figure 3 demonstrate a large discrepancy between the original ($V_{sym}$) and corrected voltage ($V_{corr\,sym}$) values for samples with NM materials of Ta or Ti. The difference can be ascribed to the stronger spin rectification effect in these samples as follows. The resistance of Ta or Ti is comparable to that of CoFeB so that a sizable amount of current flows into the CoFeB FM layer, which gives rise to a large contribution of the spin rectification effect to the dc voltage. On the other hand, for samples with less resistive NM materials than CoFeB, such as Pt or Pd, current generated by the electric field flows mostly in the NM materials. This results in a negligible spin rectification effect ($V_{corr\,sym} \sim V_{sym}$). From the resulting $V_{corr\,sym}$ values, which is closely related to the $\theta_{SH}$, we can estimate that the $\theta_{SH}$ of Ta is comparable to that of Pt, but with an opposite sign. Note that the $\theta_{SH}$ of Ti is negligible, which is consistent with weak spin-orbit coupling of this material.

In order to confirm the above procedure to obtain the pure spin pumping contribution from the FMR dc voltage, we fabricated samples of inversion stacking orders and compared them with previous samples. When the stacking order is reversed, the change in direction of the spin current induces the sign inversion of the inverse SHE and corresponding dc voltage from pure spin pumping.[23] Contrarily, the spin rectification effect is not significantly affected by the stacking order variation. Therefore, we can extract pure spin pumping and spin rectification effect by simple subtraction and addition of the results from samples with different stacking orders with the following expressions.

$$V_{pure\ SP} = \frac{1}{2}(V_{FM/NM} - V_{NM/FM}), \quad V_{spin\ rec} = \frac{1}{2}(V_{FM/NM} + V_{NM/FM}) \quad (3)$$

Figure 4 shows the experimental results of the FMR spin pumping for samples with normal (FM/NM) and inverted (NM/FM) structures, together with the calculated $V_{pure\ SP}$. Note that the stacking order variation does not affect the magnetic properties as shown in insets of Figure 4 and also resistances (less than 1% of difference) of the samples. For Pt or Pd electrodes, samples show a symmetrical sign change in dc voltage with respect to the inversion of stacking order. Therefore, the pure spin pumping contribution ($V_{pure\ SP}$) is more or less the same as the symmetric dc voltage ($V_{sym}$). However, the samples with highly resistive NM electrodes present dissimilar shapes of the dc voltage for different stacking orders. The $V_{pure\ SP}$ is observed to be a larger (or smaller) than an original symmetric dc voltage ($V_{sym}$) depending on the layer configuration, which is attributed to the larger spin rectification effect. Moreover, the $V_{pure\ SP}$ is consistent with the $V_{corr\ sym}$ values presented in Figure 3. This confirms that the spin rectification effect in those samples with highly resistive materials such as Ti or Ta can significantly modify the FMR dc voltage, which could result in over-(or under-) estimation of the spin Hall angle.

From the obtained $V_{corr\ sym}$ (or $V_{pure\ SP}$), we calculated the $\theta_{SH}$ of Ta. The spin diffusion length of Ta is around 2.5 nm, which extracted by measuring $V_{corr\ sym}$ as a function of Ta thickness, and the spin mixing conductance is $1.4\times10^{15}$/cm$^2$ from additional damping (not shown here). The rf driving field ($H_{rf}$) of 50 mW microwave power is estimated to be ~ 0.15Oe. Based on the parameters above, the spin Hall angle of Ta is calculated[12] to be around 0.05, which is smaller than the previously reported one of ~0.15.[6]

In summary, we have presented an analysis method of an FMR spin pumping signal for extraction of the spin Hall angle. Pure spin pumping contribution was obtained by comparing FMR dc voltages of FM/NM bilayers with those of a single FM layer and samples with opposite stacking orders. The relative contribution of spin rectification effect depends on the resistance ratio between FM and NM materials due to the current shunting effect through the FM layer, which could cause an error in estimation of the spin Hall angle.

This research was supported by Basic Science Research Program (NRF-2012R1A1A1041590), the framework of international cooperation program (NRF-2012K2A1A2033057) and the Pioneer Research Center Program (2011-0027908) through National Research Foundation of Korea funded by the Ministry of Science, ICT & Future Planning; the Future Semiconductor Device Technology Development Program (10044723) funded by MOTIE (Ministry of Trade, Industry & Energy) and KSRC (Korea Semiconductor Research Consortium).

**List of figures**

Figure 1: (a) Schematic of FMR spin pumping system of an FM/NM bilayer. (b) FMR spectra and (c) corresponding dc voltage vs. magnetic field (H), measured for a 10 nm CoFeB single layer. The (black) circles, (red) dotted line, and (blue) dash-dotted line represent the experimental data, fitted $V_{sym}$, and fitted $V_{asym}$ to Equation (1), respectively. The (black) solid line shows the combined fits for the CoFeB samples.

Figure 2: Voltage measured for (a) CoFeB(10)/Ta(10), (b) CoFeB(10)/Ti(10), (c) CoFeB(10)/Pd(10), and (d) CoFeB(10)/Pt(10) in nm. The (black) circles, (red) dotted line, and (blue) dash-dotted line represent the experimental data, fitted $V_{sym}$, and fitted $V_{asym}$ to Equation (1), respectively. The (black) solid line shows the combined fits for the samples.

Figure 3: The corrected dc voltages ($V_{corr\ sym}$) due solely to spin pumping for the samples with (a) Ta, (b) Ti, (c) Pd, and (d) Pt NM materials. This is calculated by subtracting the CoFeB contribution to $V_{sym}$ using Equation (2).

Figure 4: FMR spin pumping data for samples with normal (FM/NM) and inverted (NM/FM) structures for the samples with (a) Ta, (b) Ti, (c) Pd , and (d) Pt NM materials. The open squares are normal and the solid circles are in inverted stacking order. The (blue) dotted lines are the calculated $V_{pure\ SP}$ using Equation (3). The insets show the M-H hysteresis curves of the samples with different stacking order.

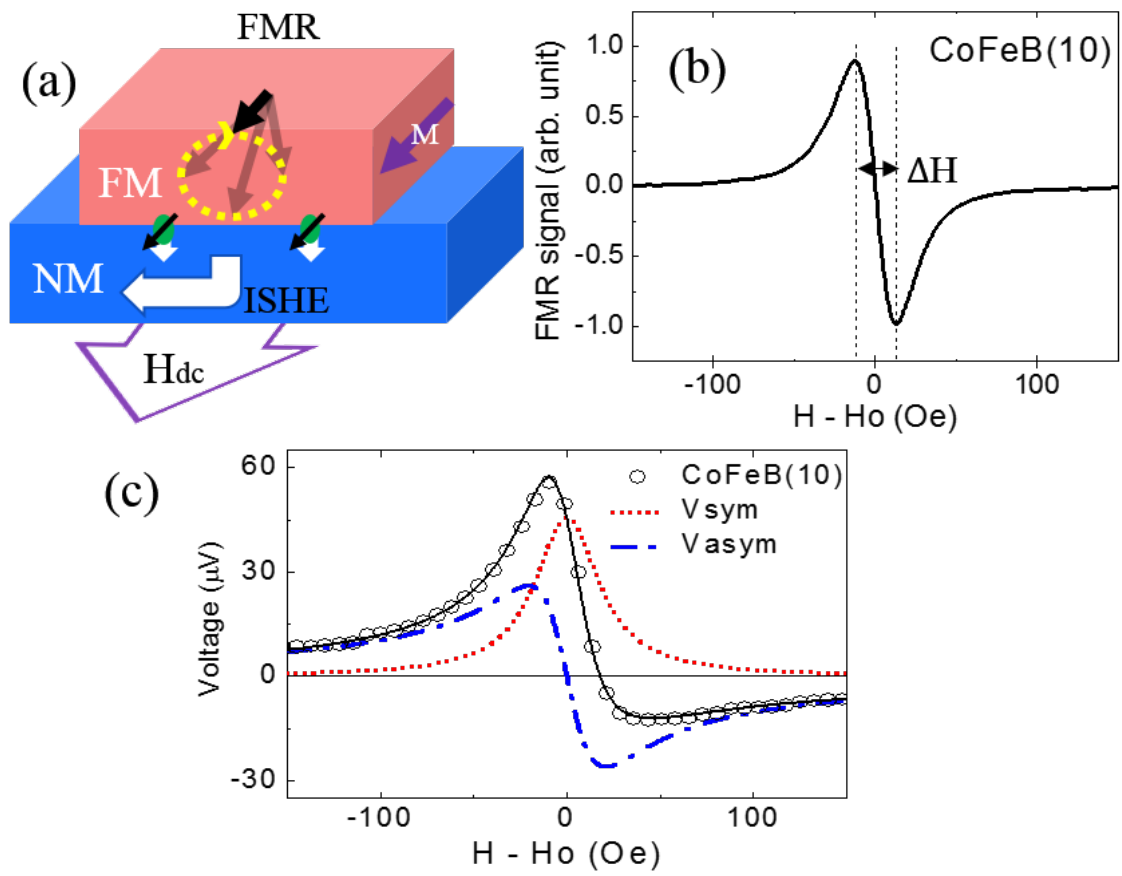

Figure 1

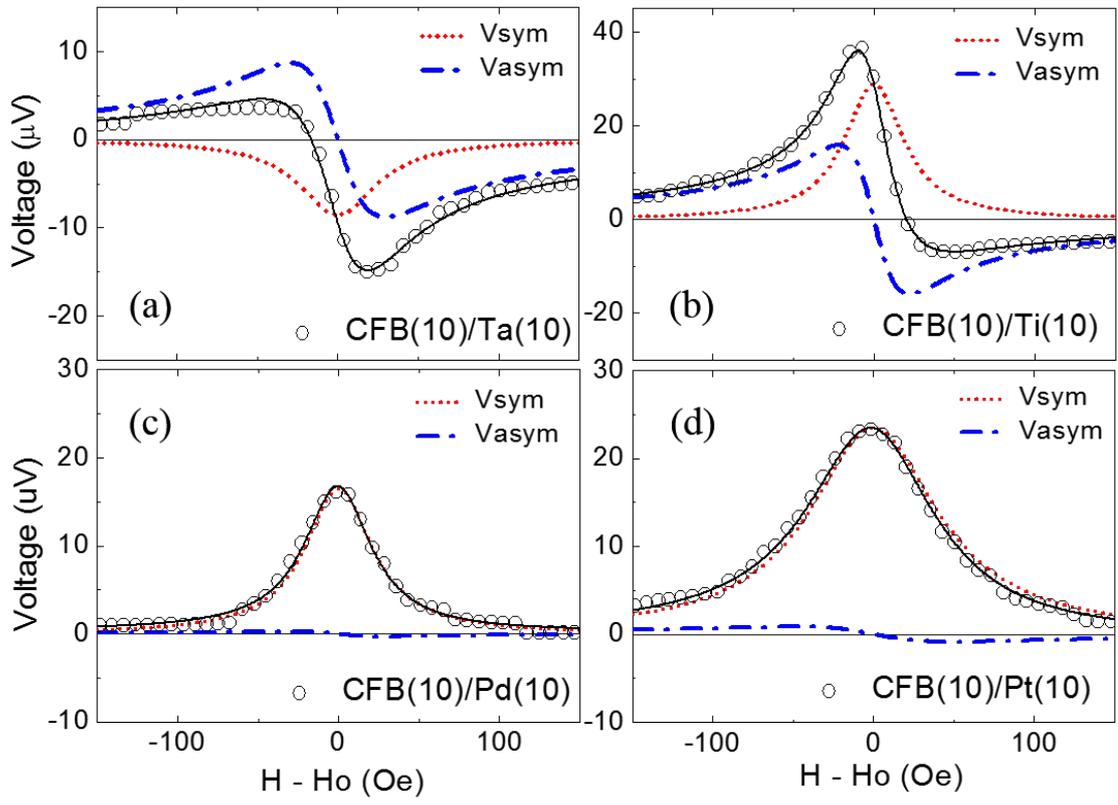

Figure 2

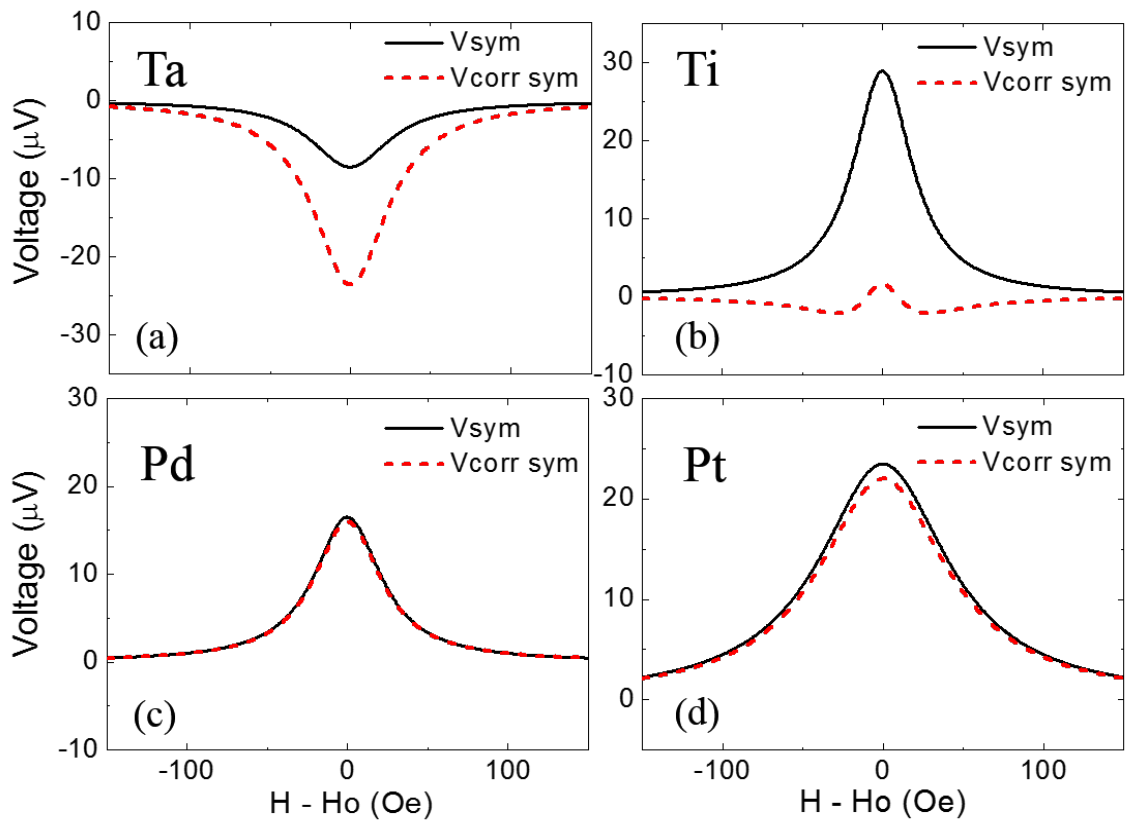

Figure 3

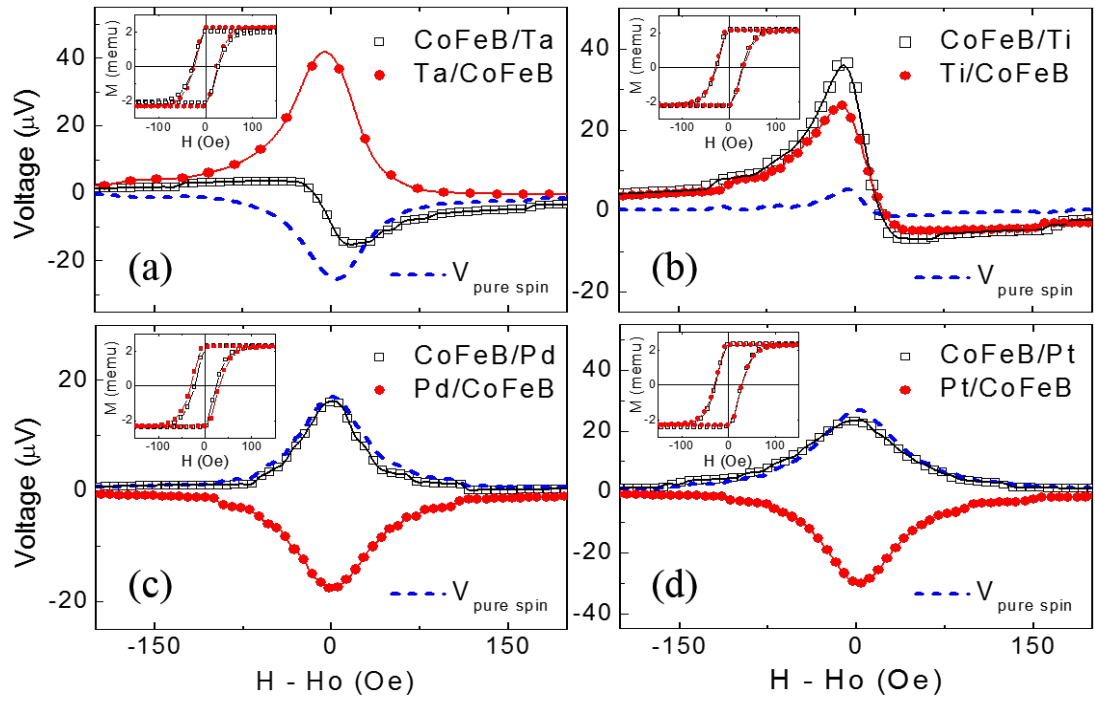

Figure 4